\documentclass[twocolumn,nofootinbib,prd]{revtex4}

\usepackage{gabe}

\begin{document}

\title{Homoclinic Orbits around Spinning Black Holes II:\par The Phase
Space Portrait}

\author{Gabe Perez-Giz${}^{**}$ and Janna Levin${}^{*,!}$}
\email{janna@astro.columbia.edu}
\email{gabe@phys.columbia.edu}
\affiliation{${}^{**}$Physics Department, Columbia University,
New York, NY 10027}
\affiliation{${}^{*}$Department of Physics and Astronomy, Barnard
College of Columbia University, 3009 Broadway, New York, NY 10027 }
\affiliation{${}^{!}$Institute for Strings, Cosmology and Astroparticle
  Physics, Columbia University, New York, NY 10027}


\begin{abstract}
In paper I in this series, we found exact expressions for the
equatorial homoclinic orbits: the separatrix between bound and
plunging, whirling and not whirling. As a companion to that physical
space study, in this paper we paint a phase space portrait of the
homoclinic orbits that includes exact expressions for the actions and
fundamental frequencies.  Additionally, we develop a reduced
Hamiltonian description of Kerr motion that allows us to track groups
of trajectories with a single global clock.  This facilitates a
variational analysis, whose stability exponents and eigenvectors could
potentially be useful for future studies of families of black hole
orbits and their associated gravitational waveforms.

\end{abstract}


\maketitle

\section{Introduction}

The
transition from inspiral to plunge is
a crucial landmark in the radiative evolution of a compact object
falling into a supermassive black hole. A natural physical divide, the
transition is also a natural conceptual divide. The inspiral
can be modeled as adiabatic evolution through a sequence of Kerr
geodesics 
\cite{Flanagan:1997sx1,glampedakis2002:2, drasco2004, drasco2005, drasco2006, lang2006}
while the plunge is currently best modeled by numerical
relativity
\cite{{pretorius2006},{herrmann2007},{campanelli2007},{campanelli2006},{baker2005},{marronetti2007},{scheel2006}}.
Inspiral gives way to plunge through an important family of separatrices.
In paper I in this series \cite{levin2008:3}, we detailed the nature
of the separatrix between bound and plunging orbits as a homoclinic
orbit -- an orbit 
in the black hole spacetime that whirls an infinite number of times
as it asymptotes to an
unstable circle.
We found exact solutions
for the family of homoclinic trajectories 
and depicted them as the infinite limit of
a sequence of zoom-whirls \cite{levin2008:3}. 
As a companion to that physical space picture, we
analyze the complementary phase space picture here. 

As discussed at some length in paper I,
formally, the homoclinic orbit lies on the intersection of
the stable and unstable manifolds of a hyperbolic invariant set. In
the black hole spacetime, the hyperbolic invariant set is recognized by
the more familiar tag ``unstable circular orbit''. To
make this connection precise from the phase space perspective, 
we examine the variational
equations -- the
equations governing the evolution of small displacements from the
circular orbits. It is straightforward to show that the energetically
bound, unstable circular orbits are hyperoblic; that is, they have an
unstable eigendirection and a stable eigendirection. We then show that
the stable and unstable eigendirections are tangent to the homoclinic
orbit in the local neighborhood of the unstable circular orbit.
In other words, 
two of the eigensolutions of the variational equations around
bound unstable circular orbits are local representations of the
homoclinic orbit. These eigensolutions capture the
qualitative and quantitative features of the separatrix discussed
in paper I, including the azimuthal motion \cite{levin2008:3}.

We begin by devising a reduced Hamiltonian formulation of equatorial Kerr
motion that natrually admits comparisons of groups of trajectories 
against a single global clock.
The variation of 
Hamilton's equations yields stability exponents for circular orbits
that could have general utility, for instance, as an estimate of
inspiral or merger timescales \cite{Cornish:2001jy,cornish2003}, or in a
coarse graining of the template space around periodic orbits \cite{levin2008}.
For completeness,
we also find explicit expressions for the actions and the frequencies

\section{Kerr homoclinic orbits in phase space}
\label{sec:phasespace}

Carter famously reduced the full geodesic equations of motion to
four first order equations in space and time coordinates \cite{carter1968}.
Despite the appeal of this accomplishment, a phase
space analysis requires variation of the full equations of motion for
both the coordinates and their conjugate momenta. For this reason we
will not work in the first-order integrated system of
equations, although we will borrow his familiar expressions. Instead, we write down a
Hamiltonian formulation of Kerr geodesic
motion and explicitly derive the equations of motion.

\subsection{Kerr Equations of Motion}
\label{subsec:eom}

Although written out in many places, including paper I
\cite{levin2008:3}, to remain self-contained we include the 
Kerr metric in Boyer-Lindquist coordinates and geometrized units
($G = c =1$): 
\begin{alignat}{1}
\label{eq:metric}
  \begin{split}
    ds^2 &= 
    - \left( 1-\frac{2Mr}{\Sigma} \right) dt^2
    - \frac{4Mar\sin^2\theta}{\Sigma} dt d\varphi \\
    &\mrph{=}
    {}+ \sin^2\theta
    \left(
    r^2+a^2 + \frac{2Ma^2r\sin^2\theta}{\Sigma}
    \right) d\varphi^2 \\
    &\mrph{=}
    {}+ \frac{\Sigma}{\Delta}dr^2
    + \Sigma d\theta^2
  \end{split}
  \quad ,
\end{alignat}
where $M, a$ denote the central black hole mass and spin angular
momentum per unit mass, respectively, and
\begin{alignat}{1}
\label{eq:DeltaSigmadefs}
  \begin{split}
    \Sigma &\equiv r^2+a^2\cos^2\theta \\
    \Delta &\equiv r^2-2Mr+a^2
  \end{split}
  \quad .
\end{alignat}
The constants of motion 
along Kerr geodesics are the rest mass of the test object, energy
$E$, axial angular momentum $L_z$, and the Carter constant $Q$
\cite{carter1968}.

In dimensionless units, 
the first-order geodesic equations are \cite{carter1968}
\begin{subequations}
\label{eq:dimcarter}
\begin{alignat}{1}
\Sigma \dot r &= \pm \sqrt{R}
\label{subeq:dimcarter-r}\\
\Sigma \dot \theta &= \pm \sqrt{\Theta}
\label{subeq:dimcarter-theta}\\
\Sigma \dot \varphi &=
\frac{a}{\Delta} \lf( 2rE - aL_z \rt)
+ \frac{L_z}{\sin^2 \theta}
\qquad \quad \elpunc{,}
\label{subeq:dimcarter-phi}\\
\Sigma \dot t&=
\frac{(r^2 + a^2)^2 E - 2arL_z}{\Delta}
- a^2 E \sin^2 \theta
\label{subeq:dimcarter-t}
\end{alignat}
\end{subequations}
where an overdot denotes differentiation with respect to the
particle's (dimensionless) proper time $\tau$ and
\begin{alignat}{1}
  \Theta(\theta) &= Q - \cos^2\theta
  \left\{
  a^2(1- E^2) + \frac{L_z^2}{\sin^2\theta}
  \right\}
  \label{eq:Thetaeq}\\
  \begin{split}
    R(r) &= -(1 - E^2)r^4 + 2r^3 - \lf[ a^2(1 - E^2) + L_z^2 \rt]r^2 \\
    &\mrph{=}  {}+ 2(aE - L_z)^2\, r - Q \Delta
    \qquad \qquad .
  \end{split}
  \label{eq:Rpoly}
\end{alignat}
The four equations ~(\ref{eq:dimcarter}), though no doubt valuable in many contexts, do not
lend themselves to a variational analysis. The formalism we will
imploy is Hamiltonian and a phase space study requires not just the
coordinates but also their conjugate momenta.
Although we start from
scratch with a Hamiltonian formulation of the dynamical equations, we
will make use of the Eqs.\ (\ref{eq:dimcarter})-(\ref{eq:Rpoly})
along the way.

As in paper I, we will restrict attention to equatorial
orbits and defer non-equatorial motion to a future work. 
Equatorial Kerr orbits have $\theta = \pi/2$, $\dot{\theta}
= 0$, and $Q=0$.

\subsection{Hamiltonian formulation}

The Hamiltonian for a relativistic non-spinning free particle of mass
$\mu$ is \cite{schmidt2002}
\begin{equation}
\label{eq:general8DHam}
  H = \half g^{\alpha\beta} p_\alpha p_\beta
  \, ,
\end{equation}
where the inverse metric components $g^{\alpha\beta}$ are functions of
the spacetime coordinates and each $p_\alpha$ is both a component of
the 4-momentum one-form and the canonical momentum conjugate to
coordinate $q^\alpha$.  

We want to build the Hamiltonian explicitly from Eq.\ ~(\ref{eq:metric}), and we
could do so just by inserting the inverse metric and turning the
crank. However, we can yield an equivalent but algebraically nicer
expression for the Hamiltonian with far less effort. To begin, consider
the terms in the Hamiltonian explicitly containing $p_r$ or $p_\theta$:
\begin{equation}
  \frac{1}{2}\left(g^{rr} p_r^2 + g^{\theta\theta} p_\theta^2\right) \quad.
\end{equation}
Since the $r, \theta$ portion of the metric $g_{\mu\nu}$ is diagonal,
that block of the inverse metric is also diagonal, with $g^{rr} =
1/g_{rr}$ and $g^{\theta\theta} = 1/g_{\theta\theta}$.  The $p_r,
p_\theta$ terms in $H$ are thus
\begin{equation}
  \frac{1}{2} \left(\frac{\Delta}{\Sigma}\right) p_r^2 +
  \frac{1}{2} \left(\frac{1}{\Sigma}\right) p_\theta^2
\end{equation}
The remaining terms in the Hamiltonian will be quadratic in the
remaining momenta $p_t$ and $p_\varphi$ with coefficients that are
functions only of $r$ and $\theta$ (since the metric, and thus the
inverse metric, are cyclic in the $t$ and $\varphi$ coordinates).  The
Hamiltonian can therefore be written as
\begin{equation}
  H(\vctr{q}, \vctr{p}) = \frac{1}{2} \left(\frac{\Delta}{\Sigma}\right)
  p_r^2 + \frac{1}{2} \left(\frac{1}{\Sigma}\right) p_\theta^2 + \frac{1}{2}F(r,
  \theta, p_t, p_\varphi) \quad,
\label{Hstep}
\end{equation}
where $F(r,\theta, p_t, p_\varphi) =F(r,\theta, E, L)$ is some
expression equivalent to $g^{tt} p_t^2 + 2g^{t\varphi} p_t p_\varphi +
g^{\varphi\varphi} p_\varphi^2$.

Notice that the $\dot r$ and $\dot \theta $ equations of (\ref{eq:dimcarter})
can be recast as
\begin{alignat}{1}
\label{eq:prpthetatrick}
  \begin{split}
    \frac{\Delta}{2\Sigma}p_r^2-\frac{R}{2\Delta\Sigma} &= 0 \\
    \frac{1}{2\Sigma}p_\theta^2-\frac{\Theta}{2\Sigma} &= 0
  \end{split}
  \quad.
\end{alignat}
Adding these equations and subtracting $1/2$ from both sides tells
us that
\begin{equation}
\frac{\Delta}{2\Sigma}p_r^2+\frac{1}{2\Sigma}p_\theta^2
  -\frac{R}{2\Delta \Sigma}
-\frac{\Theta}{
 2 \Sigma} -\frac{1}{2}= -\frac{1}{2}  \quad\quad .
\label{combo}
\end{equation}
Since $H \equiv -1/2$, the left hand side must be identical to $H$.  Matching
to Eq.\ (\ref{Hstep}), we glean that
\begin{equation}
  F(r,\theta, E, L) 
  = -\frac{R + \Delta\Theta}{\Delta \Sigma} - 1 \quad\quad ,
\label{eq:F}
\end{equation}
so that we finally get
\begin{equation}
\label{eq:Kerr8Dham}
  H =
  \frac{\Delta}{2\Sigma}p_r^2+\frac{1}{2\Sigma}p_\theta^2
  -\frac{R + \Delta\Theta}{2\Delta \Sigma} -\frac{1}{2}
  \, ,
\end{equation}
where $R$ and $\Theta$ are the functions in (\ref{eq:Rpoly}).  Note
that in dimensionless coordinates, the Hamiltonian has the same
constant value $-1/2$ along any trajectory.
We also used this form of the Hamiltonian
in Appendix A of
Ref.\ \cite{levin2008}.

Because all dependences on $E \equiv -p_t$ and $L_z \equiv p_\varphi$
are locked inside $R$ and $\Theta$ and $H$ is cyclic in $t$ and
$\varphi$, Hamilton's equations
\begin{alignat}{2}
\label{eq:Hamseqns}
  \dot{q}^\mu &= \frac{\partial H}{\partial p_\mu} \, , \quad &
  \dot{p}_\mu &= -\frac{\partial H}{\partial q^\mu}
\end{alignat}
applied to the Hamiltonian (\ref{eq:Kerr8Dham}) yield equations of
motion
\begin{widetext}
\begin{subequations}
\label{eq:8Deom}
\begin{alignat}{2}
\dot{r} &= \frac{\Delta}{\Sigma}p_{r}
\ , \quad\quad
&\dot{p}_{r}  &= 
  -\lf( \frac{\Delta}{2\Sigma} \rt)'p_{r}^{2}
  -\lf( \frac{1}{2\Sigma} \rt)'p_{\theta}^{2}
  +\lf( \frac{R + \Delta\Theta}{2\Delta\Sigma} \rt)'
  \label{subeq:prdot} \\
\dot{\theta} &= \frac{1}{\Sigma}p_{\theta}
\ , \quad\quad
&\dot{p}_{\theta} &= 
  -\lf( \frac{\Delta}{2\Sigma}\right )^{\theta}p_{r}^{2}
  -\lf( \frac{1}{2\Sigma} \rt)^{\theta}p_{\theta}^{2}
  +\lf( \frac{R + \Delta\Theta}{2\Delta\Sigma} \rt)^{\theta}
  \label{subeq:pthetadot} \\
\dot{\varphi} &=
  -\frac{1}{2\Delta\Sigma} \pd{L}{} \lf( R + \Delta\Theta \rt)
\ , \quad\quad
&\dot{p}_\varphi  &= 0
  \label{subeq:pphidot} \\
  \dot{t} &=
  \ph{+}\frac{1}{2\Delta\Sigma} \pd{E}{} \lf( R + \Delta\Theta \rt)
\ , \quad\quad
&\dot{p}_t &= 0
\label{subeq:ptdot}
\end{alignat}
\end{subequations}
\end{widetext}
where the superscripts $'$ and $\theta$ denote differentiation with
respect to $r$ and $\theta$, respectively. 
Notice, all of the Eqs.\ (\ref{eq:8Deom}) are dynamically equivalent to Eqs.\
(\ref{eq:dimcarter}).
These equations define an 8D phase space, one axis for each of
the 4 coordinates $t, r, \theta, \varphi$ and their corresponding
conjugate momenta, with $\tau$ 
parametrizing trajectories in the space.  The
Hamiltonian (\ref{eq:Kerr8Dham}) derived above governs the evolution of the
system in this 8-dimensional phase space.

A manifestly covariant form of Hamilton's
equations, equivalent to (\ref{eq:Hamseqns}), has been used in other references to deduce important
information about \emph{individual} trajectories
\cite{carter1968,schmidt2002,Hinderer:2008dm}.  We, however, want to
describe how \emph{multiple}
trajectories evolve relative to one another to locate stable and
unstable flows in phase space,
and that task requires
tracking evolution with respect to some global clock.
In the
covariant Hamiltonian picture,
the time parameter $\tau$ in (\ref{eq:Hamseqns}) flows
differently on different trajectories and is thus not a physically
viable global clock.\footnote{Mathematically, of course, $\tau$ is a
perfectly fine global clock.  After all, the Hamiltonian formalism
knows nothing about relativity and is perfectly happy to answer
physically unsensible questions like how equal $\tau$ separations
evolve with respect to ``global proper time''.}  

Coordinate time $t$
would be a good global clock, but it becomes awkward to
maintain the clock as a coordinate in the 8D phase space.
Furthermore,
all orbits move monotonically away from
the origin along the $t$ direction.\footnote{Strictly speaking, the
motion is also monotonic in the $\varphi$ direction, but topologically
identifying $\varphi = 0$ and $\varphi = 2\pi$ compactifies phase
space in the $\varphi$ direction and thus bounds the $\varphi$
motion.\cite{schmidt2002}} Consequently, no region of finite phase volume
contains any orbit in its entirety, and there are no recurrent
invariant sets.\footnote{Of course, every inidividual trajectory is
still a trivial sort of invariant set.  Since even in this space, the
phase trajectories describing the orbits in paper I
asymptote at $\tau \to \pm \infty$ to those representing unstable
circular orbits, we can still talk about their being homoclinic to an
invariant set.  Still, the language is inelegant, and having to track
the additional $t$ evolution is an unwelcome complication.} 
The 8D space, then, is not a natural backdrop
for the discussion of homoclinic orbits.

Indeed, this lack of boundedness is the hallmark of relativistic systems, in
which time itself is a coordinate. 
Luckily, we can work in a 6D space -- the phase space of spatial coordinates and their
conjugate momenta -- parameterized by coordinate time $t$. To do this
properly, we work with a new Hamiltonian function, the energy $E$,
that generates the flow parameterized by coordinate time,\footnote{Simply restricting attention to the spatial 6D
subspace of the full 8D space is not formally equivalent to using the
non-covariant Hamiltonian.  We elaborate on this in future work.}
\begin{alignat}{2}
\label{eq:6DHamseqns}
  \frac{d{q^i}}{dt} &= \frac{\partial E}{\partial p_i} \, , \quad &
  \frac{d{p_i}}{dt} &= -\frac{\partial E}{\partial q^i}
  \quad .
\end{alignat}
For details of the phase space reduction formalism see Refs.\ \cite{Lichtenberg}.
It must be stressed that we treat every $E$ in the Hamiltonian (\ref{eq:Kerr8Dham}) as
an implicit function $E(\vec{q}, \vec{p})$ of the spatial $q^i$ and
$p_i$ and solve
\begin{equation}
\label{eq:Eimplicit}
  H( \vec{q}, \vec{p}, E(\vec{q},\vec{p}) ) = -\half
\end{equation}
for $E$.\footnote{Since we consider only positive energies, we keep
the larger root in the resulting quadratic equation for $E$.}  

In other words, the spatial part of relativistic free particle motion
maps to an equivalent classical problem for which coordinate time $t$
is the time parameter and whose dynamical evolution is governed by the
Hamiltonian $E(\vec{q},\vec{p})$.  Such a space-time splitting, which
we also used in \cite{levin2008} and a fuller discussion of
which we are developing in a coming work, is dynamically exact and involves no
approximation.  The only cost is that the accumulation of proper time
$\tau$ along any trajectory (for which we will have no need in this
paper anyway) must now be tracked on the side as a separate
function.\footnote{The 6D phase space + the $\tau(\vec{q},\vec{p})$
function on that space capture the full 8D dynamics because, since $H
= -1/2$ for all trajectories, the motion is already constrained to 7D
hypersurface in the original 8D phase space.}

To get the 6D equations of motion for the Kerr system, we could
calculate $E(\vec{q},\vec{p})$ explicitly from (\ref{eq:Eimplicit}) and then
apply (\ref{eq:6DHamseqns}).  Alternately, we can realize that we have
to get the same result if we divide all the spatial equations in
(\ref{eq:8Deom}) by $\dot{t}$ (\ref{subeq:ptdot}) and immediately write
down
\begin{widetext}
\begin{subequations}
\label{eq:6Deom}
\begin{alignat}{2}
  \D{t}{r} &= \frac{1}{\dot{t}} \times
  \frac{\Delta}{\Sigma}p_{r}
\ , \quad \quad
&
    \D{t}{p_r}  &=
    \frac{1}{\dot{t}} \times
    \lf\{
    -\lf( \frac{\Delta}{2\Sigma} \rt)'p_{r}^{2}
    -\lf( \frac{1}{2\Sigma} \rt)'p_{\theta}^{2}
+\lf( \frac{R + \Delta\Theta}{2\Delta\Sigma} \rt)'
    \rt\}
  \label{subeq:dprdt} \\
 \D{t}{\theta} &= \frac{1}{\dot{t}} \times
 \frac{1}{\Sigma}p_{\theta}
\ , \quad \quad
&   \D{t}{p_\theta} &= 
   \frac{1}{\dot{t}} \times
   \lf\{
   -\lf( \frac{\Delta}{2\Sigma}\right )^{\theta}p_{r}^{2}
   -\lf( \frac{1}{2\Sigma} \rt)^{\theta}p_{\theta}^{2}
+\lf( \frac{R + \Delta\Theta}{2\Delta\Sigma} \rt)^{\theta} 
   \rt\}
 \label{eq:dpthetadt} \\
 \D{t}{\varphi} &=
 \frac{1}{\dot{t}} \times
 \lf\{
 -\frac{1}{2\Delta\Sigma} \pd{L}{} \lf( R + \Delta\Theta \rt)
 \rt\}
\ , \quad\quad
& \D{t}{p_\varphi} &= 0
 \label{subeq:dpphidt}
\end{alignat}
\end{subequations}
\end{widetext}
with the caveat that, when we calculate derivatives of Eqs.\
(\ref{eq:6Deom}), every instance of $E$ be treated as a function
$E(\vec{q},\vec{p})$ rather than as either a phase space coordinate or
a parameter.

This 6D phase space makes variational analysis straightforward:
because coordinate time $t$ is both a good global clock \emph{and} the
time parameter for (\ref{eq:6DHamseqns}), the equations dictating the
evolution in $t$ of small separations between trajectories at equal
$t$ can be derived just by linearizing Eqs.\ (\ref{eq:6Deom}).
We perform that linearization now.

\subsection{The variational equations}
\label{sec:KLformalism}

We work exclusively in the 6D phase space and introduce
the following notational simplification.  Because the distinction
between $q$'s and $p$'s as components of vectors and one-forms,
respectively, has to do with their behavior in the 4D manifold of the
Kerr spacetime and not with their function in the phase space, where
they are merely coordinates labeling points, we will henceforth drop
the superscript/subscript distinction.  Instead, we will refer to both
$q^i$ and $p_i$ as components $X_i$ (with a subscript) of a single
six-dimensional coordinate vector
\begin{equation}
\label{eq:Xdef}
  \vctr{X} \equiv
  \sixvec{r}{p_r}{\theta}{p_\theta}{\varphi}{p_\varphi}
  \quad .
\end{equation}
This allows us to write Hamilton's equations in the compact form
\begin{equation}
\label{eq:Xeom}
\frac{d\vctr{X}}{dt}={\vctr{f}}(\vctr{X})
\quad ,
\end{equation}
where the components of $\vctr{f}$ can be read off Eq.\
(\ref{eq:6Deom}).

Now consider an arbitrary reference trajectory $\vctr{X}(t)$ in phase
space and the vector $\vctr{\delta X}(t)$ of small displacements from
points on $\vctr{X}(t)$ to points at the same coordinate time on
neighboring phase trajectories.  The first order equations of motion
for $\vctr{\delta X}(t)$ are the linearized full equations of motion
(\ref{eq:Xeom}) around $\vctr{X}(t)$.  Specifically,
\begin{equation}
\label{eq:lineom}
\begin{split}
  \D{t}{\,\vctr{\delta X}(t)} &=
  \pd{\vctr{X}}{\vctr{f}}\evalat{\vctr{X}(t)} \vctr{\delta X}(t)\\
  &\equiv \mtrx{K}(\vctr{X}(t))\, \vctr{\delta X}(t)
\end{split}
  \quad ,
\end{equation}
or, componentwise,
\begin{alignat}{2}
  \D{t}{\,\delta X_i(t)} &= K_{ij}(\vctr{X}(t))\, \delta X_j(t)
  \label{eq:lineom-comp}\\
  \begin{split}
    K_{ij}(\vctr{X}) &\equiv \pd{X_j}{f_i}\evalat{\vctr{X}}
    \qquad \quad , \\
    &= \pd{X_j}{f_i}\evalat{\substack{\text{fixed}\\E}}
    + \pd{X_j}{E} \pd{E}{f_i}
  \end{split}
  \label{eq:Kcomps}
\end{alignat}
where the last equality stems from the caveat reagarding equations
(\ref{eq:6Deom}).

Equation (\ref{eq:lineom}) is a system of first-order linear ordinary
differential equations whose coefficients $K_{ij}(t)$ depend
implicitly on time through the solutions $\vctr{X}(t)$ to
(\ref{eq:Xeom}).  The solution to such a system can always be
expressed in terms of a fundamental matrix \cite{Boyce} $\mtrx{L}(t;
\vctr{X}_0)$ that depends on the point $\vctr{X}_0$ on the reference
trajectory at which we define the initial displacement vector
$\vctr{\delta X}_0$ and that satisfies
\begin{align}
\label{eq:delXoft}
    \vctr{\delta X}(t) &= \mtrx{L}(t;\vctr{X}_0)\,\vctr{\delta X}_0
\quad ,
\end{align}
where $\mtrx{L}(t=0;\vctr{X}_0)$ is the identity matrix.

The goal of variational analysis is to find $\mtrx{L}$, which we can
equivalently think of as the time evolution operator for small
displacements.  Given the equations of motion (\ref{eq:Xeom}), we can
always calculate the matrix $\mtrx{K}$, but in general there is no
corresponding analytic expression for $\mtrx{L}$.  However, 
$\mtrx{K}$ on
equatorial circular orbits is the constant matrix\footnote{Although
Eq.\ (\ref{eq:Kijcirc}) can be expressed solely in terms of the black
hole spin $a$ and the constant radial coordinate $r$ of the circular
orbit, we have left it in this form for readability.}
\begin{equation}
\label{eq:Kijcirc}
\mtrx{K} = \frac{1}{\gamma\Sigma}
\left(
\begin{array}{cccccc}
0 & \Delta  & 0 & 0 & 0 & 0\\ 
\frac{R^{''}}{2 \Delta}  & 0 & 0 & 0 & 0 &
\pm \frac{2 r^{3/2}}{\gamma\Delta}\\
0 & 0 & 0 & 1 & 0 & 0\\
0 & 0 & \frac{\Theta^{\theta\theta}}{2} & 0 & 0 &0 \\
\mp \frac{2  r^{3/2}}{\gamma\Delta} & 0 & 0 & 0 & 0 &
\frac{ r^2}{\gamma^2\Delta} \\
0 & 0 & 0 & 0 & 0 & 0
\end{array}
\right)
\quad ,
\end{equation}
where $R''$ and $\Theta^{\theta\theta}$ are the second derivatives
with respect to their arguments of $R(r)$ and $\Theta(\theta)$,
respectively, and $\gamma$ is a shorthand for
\begin{equation}
\left.\gamma \equiv \dot{t}(r)\right |_{r=\text{ circular \ orbit}}
\quad .
\end{equation}

Since $\mtrx{K}$ is constant, $\mtrx{L}$ has the
form\footnote{Considerable analytic insight into $\mtrx{L}$ is also
possible when the $\mtrx{K}(t)$ is \emph{periodic} in time $t$, a
situation that arises when the reference trajectory $\vctr{X}(t)$ is
itself periodic and which we tackle for Kerr orbits in a future work.}
\begin{equation}
\label{eq:LforconstK}
  \mtrx{L}(t) = e^{\mtrx{K} t}
\end{equation}
and shares its eigenvectors with $\mtrx{K}$.  Finding the
eigensolutions of (\ref{eq:lineom}) is therefore tantamount to finding
eigenvalues and eigenvectors of $\mtrx{K}$.

\subsection{Eigensolutions of the variational equations}
\label{sec:evecs}

\begin{figure}[hb]
  \vspace{-5pt}
  \centering
  \includegraphics[width=0.58\textwidth]{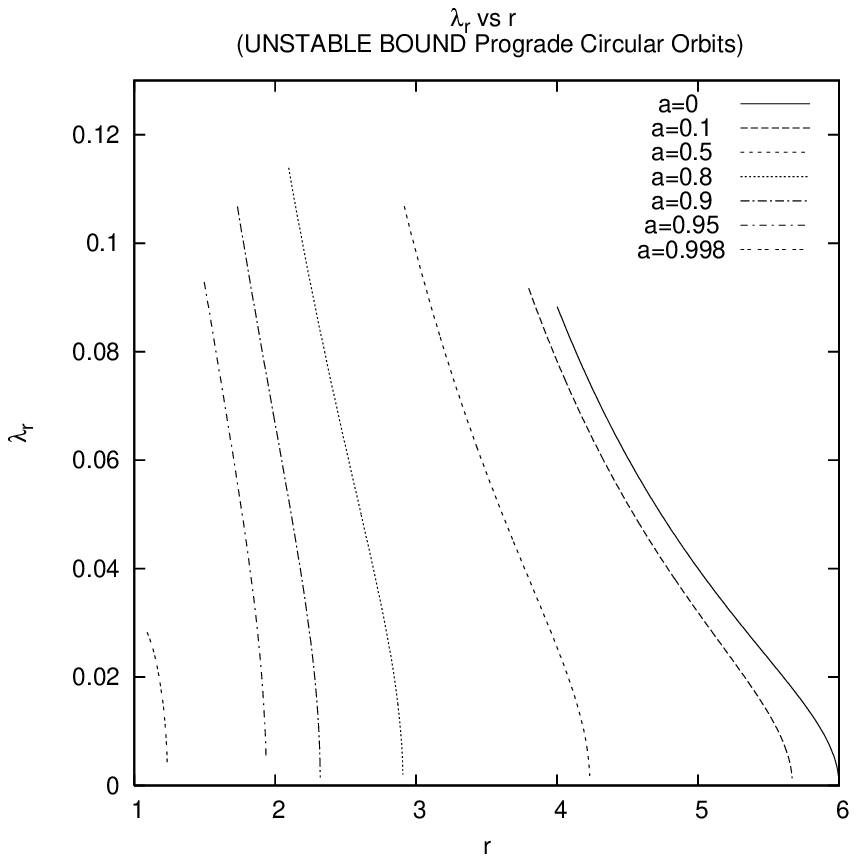}
  \includegraphics[width=0.58\textwidth]{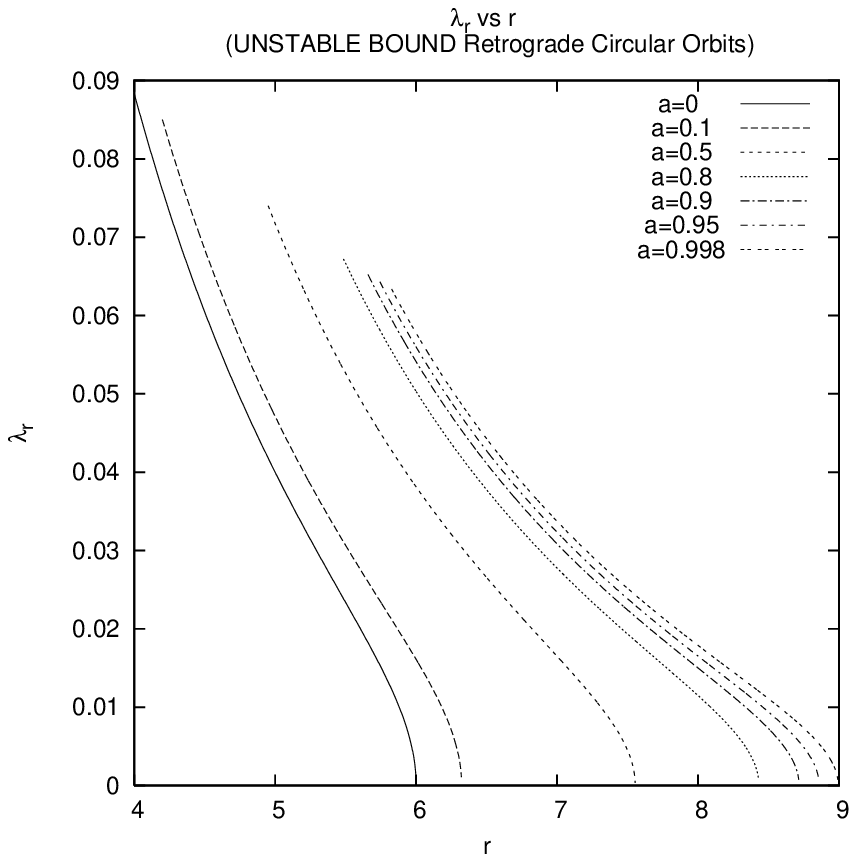}\\
\hfill
\vspace{-10pt}
  \caption{The dimensionless real-valued stability exponent
  $\lambda_r$ (measured in units of $M^{-1}$) for unstable circular
  orbits with $E < 1$ for various spins $a$.  Left: Prograde
  orbits. Right: Retrograde orbits.}
  \label{fig:lamrbd}
\end{figure}

The eigenvalues $\lambda$ of $\mtrx{K}$ are solutions to
\begin{equation}
\label{det}
\left | \mtrx{K} -\lambda \mtrx{I} \right |=0
\end{equation}
and come in 3 pairs of equal and opposite eigenvalues whose magnitude we denote as
\begin{alignat}{3}
\label{eq:evals}
   \lambda_r &=
   \frac{1}{\gamma\Sigma} \sqrt{\frac{R''}{2}}\quad, &\quad
   \lambda_\theta &=
   \frac{1}{\gamma\Sigma}
  \sqrt{\frac{\Theta^{\theta\theta}}{2}}\quad, &\quad
   \lambda_\varphi &= 0 \quad .
\end{alignat}
(See also \cite{pretorius2006})
The eigensolutions associated with the $\lambda_\theta$ and $\lambda_\varphi=0$
eigenvalues are extremely revealing in their own right.  Presently,
however, our concern is the eigensolutions associated with
$\lambda_r$, and we defer a complete discussion of the eigenvectors
of $\mtrx{K}$ to a future work.

The $\lambda_r$ may be real or imaginary depending on the sign of
\begin{align}
\label{eq:Rppcirc}
    \frac{R''}{2} &= 12r \left[ 1 - (1 - E^2)r \right] 
     - 2 \left[ a^2(1-E^2) + L_z^2 \right] \nonumber \\
    &= -\frac{r^{1/2} (r^2 - 6r \pm 8ar^{1/2} - 3a^2)}
    {r^{3/2} - 3r^{1/2} \pm 2a}
  \quad \elpunc{,}
\end{align}
where we have used the $(E,L_z)$ found in Ref.\ \cite{Bardeen1972} and
used in paper I \cite{levin2008:3} to write $R''$ in terms of $r$
alone. The plus/minus signs indicate prograde/retrograde. 
On the unstable circular orbits of interest to us
($r_{\text{ibco}} < r < r_{\text{isco}}$), $R''$ is positive and
$\lambda_r$ is real and plotted as a function of $r$ for various
values of $a$ in Fig.\ \ref{fig:lamrbd}.

The (unnormalized) eigenvectors
\begin{align}
  \label{eq:urplus}
  \vctr{u}^{\sss{(u)}}_r &=
  \begin{pmatrix}
    \Delta\,, & \ph{+}\sqrt{\frac{R''}{2}}\,, & 0\,, & 0\,, &
    \mp \frac{2 r^{3/2}}{\gamma\sqrt{R''/2}}\,, & 0
  \end{pmatrix}^T \\
  \label{eq:urminus}
  \vctr{u}^{\sss{(s)}}_r &=
  \begin{pmatrix}
    \Delta\,, & -\sqrt{\frac{R''}{2}}\,, & 0\,, & 0\,, &
    \pm \frac{2 r^{3/2}}{\gamma\sqrt{R''/2}}\,, & 0
  \end{pmatrix}^T
\end{align}
associated with $\pm\lambda_r$ are also real.  Combining
(\ref{eq:delXoft}) and (\ref{eq:LforconstK}), each
eigenvalue/eigenvector pair yields a corresponding eigensolution
\begin{alignat}{1}
\label{eq:unstabeigendeltas}
  \begin{split}
    \vctr{\delta X^{\sss{(u)}}_r}(t) &=
    c^{\sss{(u)}} e^{+\lambda_r t} \vctr{u}^{\sss{(u)}}_r \\
    \vctr{\delta X^{\sss{(s)}}_r}(t) &=
    c^{\sss{(s)}} e^{-\lambda_r t} \vctr{u}^{\sss{(s)}}_r
  \end{split}
\end{alignat}
to the variational equation (\ref{eq:lineom}), where the constants
$c^{\sss{(u,s)}}$ reflect where we choose to set $t=0$.

\subsection{Relation to the homoclinic orbits}
\label{sec:homoevecs}

We now build the case that in the neighborhood of
$\vctr{X^{\text{circ}}}(t)$, the linearized solutions
$\vctr{X^{\sss{(u,s)}}}(t)$ coincide with exact homoclinic solutions
$\vctr{X^{\text{hc}}}(t)$.  For simplicity, we focus first on the
unstable solution in (\ref{eq:unstabeigendeltas}), which corresponds to a
linearized solution
\begin{equation}
\label{eq:unstablinplussoln}
  \vctr{X^{\sss{(u)}}}(t) =
  \vctr{X^{\text{circ}}}(t) + \vctr{\delta X^{\sss{(u)}}_r}(t)
\end{equation}
to the full equations of motion (\ref{eq:Xeom}).

Some of the similarities between the linearized and homoclinic orbit
are self-evident.  The absence of $\theta$ and $p_\theta$ components
in $\vctr{X^{\sss{(u)}}}(t)$ indicates that the orbit remains
equatorial, and the identical signs on the $r$ and $p_r$ components
reflect the fact that small displacements from the circular orbit
along the eigendirection run away exponentially to larger radial
positions and velocities on an e-folding timescale $\lambda_r^{-1}$.
The absence of a $p_\varphi$ component in $\vctr{\delta
X^{\sss{(u)}}_r}(t)$ indicates that the linearized orbit has the same
angular momentum $L_z$ as $\vctr{X^{\text{circ}}}(t)$.

Less self-evident is the fact that, like the homoclinic orbit, the
linearized orbit also has the same energy $E$ as the circular orbit.
To see this, note that since the Hamiltonian $E = E(\vctr{X})$ is a
function of the phase space coordinates, the energy difference $\delta
E = E^{\text{circ}} - E^{\text{lin}}$ can be expanded as a power
series in the components of $\vctr{\delta X^{\sss{(u)}}_r}$.  Because
the derivatives of all phase variables except $\varphi$ vanish on the
circular orbit and $\delta p_\varphi = 0$, the first order
contribution to that expansion vanishes,
\begin{equation}
\label{eq:deltaE1}
  \begin{split}
    {\delta E}^{(1)} &= 
    \pd{x^i}{H_{6D}}\evalat{r_u} \delta x^i +
    \pd{p_i}{H_{6D}}\evalat{r_u} \delta p_i \\
    &= -\D{t}{p_i}\evalat{r_u} \delta x^i +
    \pd{t}{x^i}\evalat{r_u} \delta p_i\\
    &= \D{t}{\varphi} \delta p_\varphi = 0
  \end{split}
  \quad .
\end{equation}
The second order variation in the energy becomes
\begin{widetext}
\begin{equation}
\label{eq:deltaE2}
  \begin{split}
    {\delta E}^{(2)} 
&= 
    \pdd{x^i}{x^j}{H_{6D}}\evalat{r_u} \delta x^i \delta x^j +
    \pdd{p_i}{p_j}{H_{6D}}\evalat{r_u} \delta p_i \delta p_j +
    2\pdd{x^i}{p_j}{H_{6D}}\evalat{r_u} \delta x^i \delta p_j
    \nonumber \\
    &=
    -\pd{x^i}{}\D{t}{p_j}\evalat{r_u} \delta x^i \delta x^j +
    \pd{p_i}{}\D{t}{x^j}\evalat{r_u} \delta p_i \delta p_j +
    2\pd{x^i}{}\D{t}{x^j}\evalat{r_u} \delta x^i \delta p_j \nonumber \\
    &=
    -\pd{r}{}\D{t}{p_r}\evalat{r_u} \delta r^2 +
    \pd{p_r}{}\D{t}{r}\evalat{r_u} \delta p_r^2 +
    2\pd{r}{}\D{t}{r}\evalat{r_u} \delta r \delta p_r \nonumber \\
    &=
    -K_{p_r r}\evalat{r_u} \delta r^2 +
    K_{r p_r}\evalat{r_u} \delta p_r^2 +
    2K_{rr}\evalat{r_u} \delta r \delta p_r
  \end{split}
  \quad .
\end{equation}
\end{widetext}
Using Eq.\ (\ref{eq:Kijcirc}) and the fact that
\begin{equation}
\label{eq:eigendelprdelr}
  \delta p_r = \frac{1}{\Delta} \sqrt{\frac{R''}{2}} \delta r
\end{equation}
on the eigensolution, we find that
\begin{equation}
\label{eq:deltaE2zero}
  \begin{split}
&    {\delta E}^{(2)} \nonumber \\
&= \frac{ \delta r^2}{\gamma\Sigma}
    \lf(
    -K_{p_r r}\evalat{r_u} +
    K_{r p_r}\evalat{r_u}  \frac{R''}{2\Delta^2} +
    2K_{rr}\evalat{r_u} \sqrt{\frac{R''}{2\Delta^2}}
    \rt) \nonumber \\
    &= \frac{ \delta r^2}{\gamma\Sigma} 
    \lf(
    \frac{R''}{2\Delta} +
    \Delta \lf(\frac{1}{\Delta^2}\rt) \frac{R''}{2} + 0
    \rt) \nonumber \\
    &= 0
  \end{split}
  \quad .
\end{equation}
A similar result holds for $\vctr{X^{\sss{(s)}}}(t)$, despite the
addition of an overall minus sign in (\ref{eq:eigendelprdelr}), since
through second order $\delta E$ depends on $\delta p_r^2$.  Continuing
this process to higher orders is beyond the algebraic patience of the
authors, but at least through second order in the variations, the
linearized solutions describe orbits with the same $E$ and $L$ as the
unstable circular orbit.

The $\varphi$ component of $\vctr{\delta X^{\sss{(u)}}_r}(t)$ merits
more discussion.  The ratio $\delta\varphi/\delta r$ is fixed, so that
$\delta\varphi$ does not merely represent an arbitrary overall
translation in $\varphi$.
Instead, this component indicates how the phasing difference between
the linearized orbit and the circular orbit changes as the radial
separation between the two orbits grows.  Notice also that since
$\vctr{\delta X^{\sss{(u)}}_r}(t) \to 0$ as $t \to -\infty$ regardless
of how $c^{\sss{(u)}}$ is chosen, the linearized solution describes an
orbit that is in phase with the circular orbit in the infinite past.
As discussed in paper I \cite{levin2008:3}, there is a unique choice of phase
for a homoclinic orbit that will synchronize it with the
circular orbit in the infinite past. Apparently, the linearized
eigensolution goes so far as to select the \emph{phase} of the
homoclinic orbit it locally approximates.\footnote{Of course we can
have a homoclinic orbit of any phase still line up with the linearized
solution simply by adding an overall $\varphi$ shift to $\vctr{\delta X^{\sss{(u)}}_r}(t)$.}
The import is that the linearization captures detailed information
about neighboring orbits, including phase information.

Analogously, the linearized solution
\begin{equation}
\label{eq:unstablinminussoln}
  \vctr{X^{\sss{(s)}}}(t) =
  \vctr{X^{\text{circ}}}(t) + \vctr{\delta X^{\sss{(s)}}_r}(t)
\end{equation}
synchronizes with the circular orbit at $t = +\infty$.  We can now
understand the signs of the $\delta\varphi$ components of both
eigenvectors.  In $\vctr{\delta X^{\sss{(u)}}_r}(t)$ it has the
opposite sign as $\delta r$ because as the displaced orbit moves to
larger $r$, its $\txtD{t}{\varphi}$ drops, and it \emph{lags} the
circular orbit with which it was synchronized at $t = -\infty$.  In
$\vctr{\delta X^{\sss{(s)}}_r}(t)$, in contrast, $\delta\varphi$ and
$\delta r$ have the same sign: since the circular orbit will
accumulate azimuth faster than the displaced orbit as it spirals in,
it must begin \emph{ahead} of the circular orbit in phase if the two
are to synchronize at $t = +\infty$.

\begin{figure*}
  \centering
\includegraphics[width=0.45\textwidth]{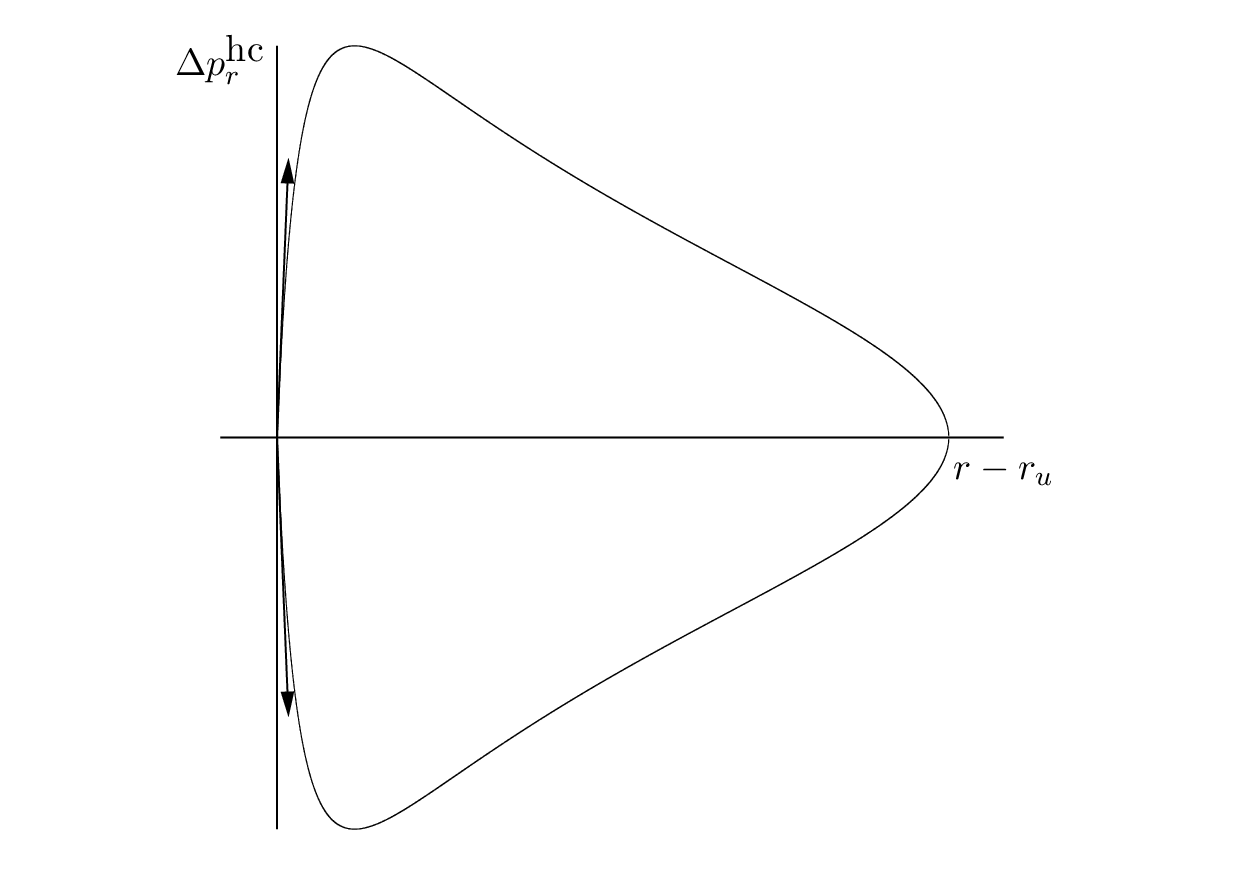}
  \includegraphics[width=0.45\textwidth]{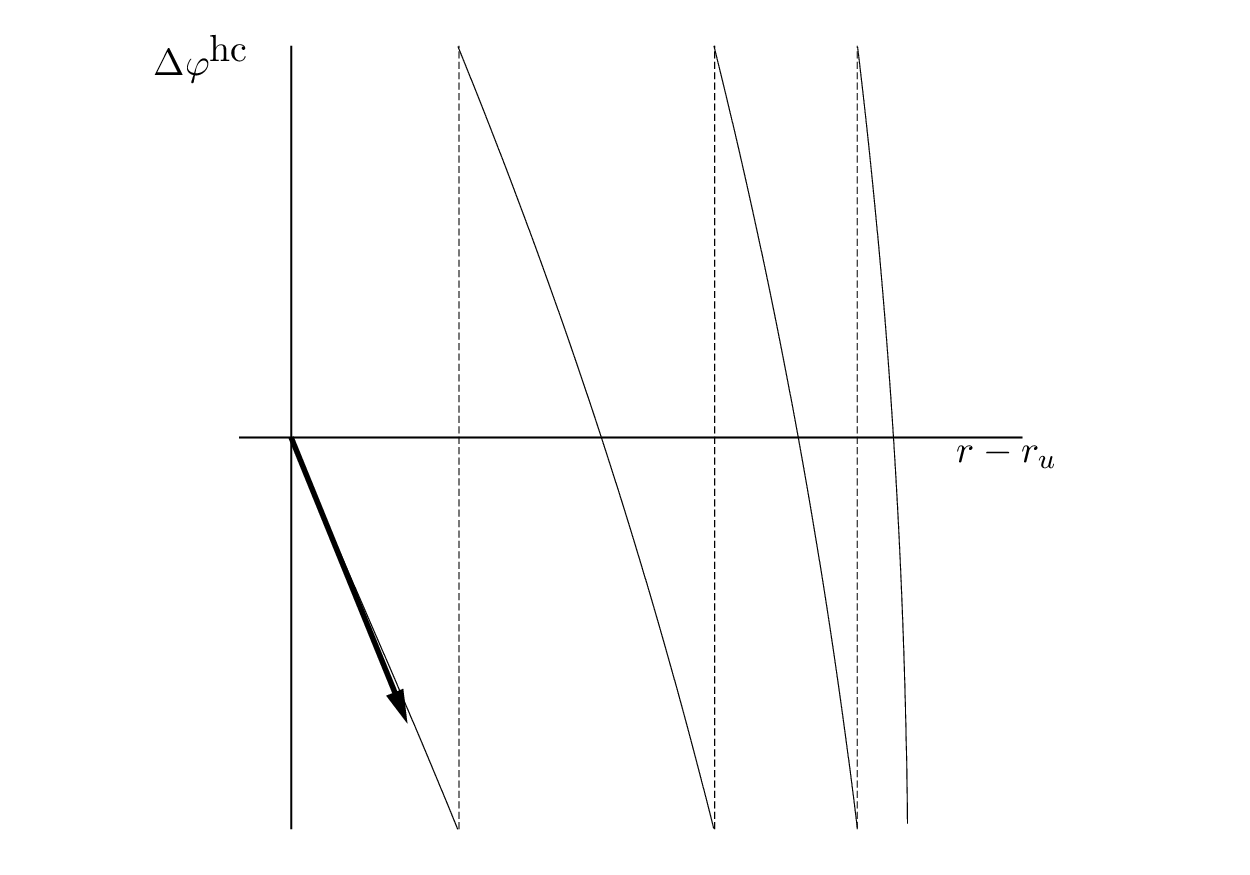}
\hfill
  \caption{Projections of the eigenvector $\vctr{u}^{\sss{(u)}}_r$, to
  which the linearized separation $\vctr{\delta X}^{\sss{(u)}}$ is
  proportional, overlayed with the actual coordinate differences
  $\vctr{X}^{\text{hc}} - \vctr{X}^{\text{circ}}$ in the phase space.
  In the $\Delta\varphi$ plot, we have identified $-\pi$ at the bottom
  of the plot and $\pi$ at the top.  The plots, intended to be
  schematic, are around an unstable circular orbit at $r_u = 2.2$ for
  $a=0.8$.}
  \label{fig:homoevectangents}
\end{figure*}

Now, as discussed in paper I \cite{levin2008:3}, the
two linearized solutions $\vctr{X^{\sss{(u)}}}(t)$ and
$\vctr{X^{\sss{(s)}}}(t)$ do not coincide with the same
homoclinic orbit, but rather with two homoclinic orbits that differ by
a phase. Since circular orbits that differ by a phase belong to the
same invariant set, we continue to refer to
these as homoclinic and not heteroclinic trajectories.

\subsection{Phase portraits}
\label{sec:phaseportraits}
To make the coincidence between the linearized solutions and the
homoclinic orbits manifest, we examine a phase portrait of the
homoclinic orbit and the linearized solutions.  Again, we use the
radial coordinate $r$ along the homoclinic orbit as our global time
parameter.  The required expression for $p_r$ in terms of $r$ for the
homoclinic orbit follows from Eqs.\ ~(\ref{subeq:prdot}).
The result is
\begin{equation}
\label{eq:prhomo}
  p_r(r) = \frac{\sqrt{R(r)}}{\Delta} 
\end{equation}
for outbound motion and the negative of the same expression for
inbound motion.  Together with the exact solutions from paper I \cite{levin2008:3},
(\ref{eq:prhomo}) generates the exact phase curves of the homoclinic
orbit.  Fig.\ \ref{fig:homoevectangents} overlays a homoclinic orbit
and the corresponding linearized orbit $\vctr{X^{\sss(u)}}$.  By
construction, the orbits are coincident at $t = -\infty$.

For illustration, we have plotted the case $a=0.8$ with an associated
unstable circular orbit at $r_u = 2.500536$.  Since both orbits are
equatorial (so that $\theta$ motion can be suppressed) and have the
same $L_z$, a 3D orbit in $r, p_r, \varphi$ space captures all the
dynamical information, and each panel of Fig.\
\ref{fig:homoevectangents} shows the projections of the two orbits
into a plane.  The curves in Fig.\ \ref{fig:homoevectangents} are the
coordinate separations between the homoclinic and circular orbits,
with the various projections of the separation eigenvectors overlayed.
They confirm the claim made in paper I \cite{levin2008:3} that the
global stable and unstable manifolds of the circular orbits are
tangent at the circular orbits to the local stable and unstable
manifolds defined by the eigensolutions to the variational equations.

\subsection{Action-angle variables}
\label{sec:actionangle}

In an action-angle formulation
\cite{schmidt2002,Glampedakis:2005hs,Lichtenberg} of Kerr motion, the
Hamiltonian is reformulated in terms of constant momenta $J_i$ called
actions and canonically conjugate angle variables $\psi_i$ that
increase linearly with time at rates $\omega_i$.  Fourier expansions
of orbit functionals in terms of the fundamental frequencies
$\omega_i$ are the basis of frequency-domain radiative evolution
codes, and Ref.\ \cite{Hinderer:2008dm} develops a description of the
inspiral dynamics entirely in terms of action-angle variables.  For
completeness, we include exact expressions for the frequencies and
actions of homoclinic orbits.

\subsubsection{Fundamental frequencies}
Because the equatorial Kerr system is two dimensional and integrable,
every bound orbit has an associated pair of fundamental
frequencies\footnote{Even equatorial orbits have a third frequency
$\omega_\theta$ associated with small oscillations about the
equatorial plane.  We discuss the significance of these frequencies
for all equatorial orbits in a separate work.} 
\begin{subequations}
\label{eq:omegadefs}
\begin{alignat}{1}
  \omega_r &\equiv \frac{2\pi}{T_r}
  \label{eq:omegardef}
  \\
  \begin{split}
    \omega_\varphi &\equiv 
    \frac{1}{T_r} \int_{0}^{T_r} \absval{\D{t}{\varphi}} dt
    \\
    &= \frac{2 \int_{r_p}^{r_a} dr\, \absval{\D{r}{\varphi}}}
    {\int_{r_p}^{r_a} dr\, \D{r}{t}}
  \end{split}
  \label{eq:omegaphidef}
  \quad .
\end{alignat}
\end{subequations}
Because their radial period is infinite, $\omega_r = 0$ for homoclinic
orbits.  Homoclinic orbits also whirl an infinite amount as they
approach their periastron $r_u$, so both the numerator and denominator
of (\ref{eq:omegaphidef}) diverge.

However, as we show in paper I \cite{levin2008:3}, the divergences in
both $T_r$ and the accumulated azimuth $\varphi$ can be traced to
specific terms of the form
\begin{equation}
\label{eq:phitdivergent}
  \lf.
  \begin{aligned}
    \varphi &\to 2 \frac{\Omega_u}{\lambda_r}
    \atanh{ \sqrt{\frac{r_u}{r} \frac{r_a - r}{r_a - r_u}} }
    \\
    t &\to 2 \frac{1}{\lambda_r}
    \atanh{ \sqrt{\frac{r_u}{r} \frac{r_a - r}{r_a - r_u}} }
  \end{aligned}
  \rt.
  \quad
  \text{as $t \to T_r = \infty, r \to r_u$ .}
\end{equation}
Their ratio thus converges to $\Omega_u \equiv
\txtD{t}{\varphi}(r_u)$, the constant coordinate velocity of the
circular orbit at $r_u$.

The azimuthal frequency for the homoclinic orbit and its
associated unstable circular orbit are thus the same,
\begin{equation}
\label{eq:omegashc}
  \begin{split}
    \omega_r^{\text{hc}} &= 0\,, \\
    \omega_\varphi^{\text{hc}} &= \absval{\Omega_u} =
    \frac{1}{r_u^{3/2} + a^2}
  \end{split}
  \, .
\end{equation}
That allows us to make a nice statement: the stable and unstable
circular orbits determine the lower and upper bounds, respectively of
the $\omega_\varphi$'s of all eccentric bound orbits with a given
$L_{\text{isco}} < L_z < L_{\text{ibco}}$.

\subsubsection{Actions}
Each action $J_i$ of a bound orbit is defined by
\begin{equation}
\label{eq:Jgendefs}
  J_i \equiv \oint p_i\, dq_i
\end{equation}
where the integral is taken over the projection of the orbit into the
$q_i, p_i$ plane.  Since $p_\varphi = L_z$ is constant, $J_\varphi =
2\pi L_z$ for any orbit.  The radial action $J_r$ is the area enclosed
by closed $(r,p_r)$ curves like that of Fig.\
\ref{fig:homoevectangents},
\begin{equation}
\label{Jrdef}
  J_r \equiv \oint p_r(r) \, dr = 
  2 \int_{r_p}^{r_a} dr 
  \quad .
\end{equation}
For arbitrary orbits, (\ref{Jrdef}) at best reduces to elliptic
integrals, but for the homoclinic orbit, $J_r$ can be written as an
exact function of $r_u$ alone.  The result, derived in the Appendix,
is
\begin{widetext}
\begin{equation}
  \begin{split}
  J_r^{\text{hc}} &= 2 \sqrt{1 - E^2} \times
  \Biggl\{
  -\sqrt{r_u \lf(r_a - r_u \rt)}
  + 2 \frac{2E^2 - 1}{1 - E^2}
  \atan \sqrt{ \frac{  r_a - r_u  }{r_u} }
  \Biggr\}
  \\
  &\mrph{=}
  + \frac{2}{\sqrt{1 - a^2}}
  \Biggl\{
  \sqrt{R(r_{\sss{-}})}
  \atanh \sqrt
    { 
      \frac{r_{\sss{-}}}{r_a - r_{\sss{-}}}
      \frac{r_a - r_u}{r_u}
    }
  -\sqrt{R(r_{\sss{+}})}
  \atanh \sqrt
    { 
      \frac{r_{\sss{+}}}{r_a - r_{\sss{+}}}
      \frac{r_a - r_u}{r_u}
    }
  \Biggr\}
  \end{split}
  \quad .
\end{equation}
\end{widetext}

\section{conclusion}

Although the results of this paper are self-contained, the phase space
portrait is a direct complement to the physical space portrait of
paper I \cite{levin2008:3}. Both approaches identify the separatrix
between bound and plunging orbits with a homoclinic trajectory that
whirls an infinite number of times on asymptotic approach to a circle.

Although the intention was to detail a profile of the separatrix, the
technical results of this paper 
could have further
utility. In partcular, the whirling stages of
trajectories in the vicinity of the homoclinic set might be modeled as
variations around the circular orbit using the eigenvectors and
eigenvalues found here. In the future, we aim to generalize this approach to
capture orbits around the periodic set \cite{levin2008} and to move out of the
equatorial plane \cite{levin2008:2,grossman2008}.

\begin{figure}
  \centering
\includegraphics[width=0.25\textwidth]{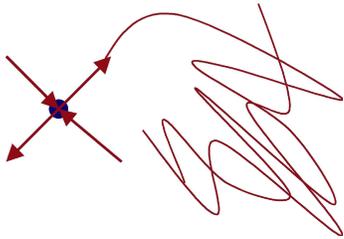}
\hfill
  \caption{Schematic of a homoclinic tangle.  The curve above
  represents the repeated intersection of a homoclinic orbit of the
  perturbed system with the $r,p_r$ phase plane.  The large dot
  represents the intersection of a circular orbit in the perturbed
  system along with the eigenvectors denoting its local stable and
  unstable manifolds.}
  \label{fig:tangle}
\end{figure}

Another connection that should be made in a dynamical discussion of
the separatirx is its role as
the divide between chaotic and non-chaotic behavior. 
The geodesic motion of a non-spinning test particle around a Kerr
black hole is known to be integrable \cite{carter1968}. There are as many constants of
motion as there are canoncial momenta in this Hamiltonian system and
the motion can therefore be confined to regular tori in an
action-angle set of coordinates. 

However, the presence of a homoclinic orbit
indicates the Kerr system is vulnerable to chaos
\cite{bombelli1992,suzuki1997,Suzuki:1999si,Kiuchi:2004bv}. Under
perturbation, the stable and unstable manifolds that previously
coincided along the homoclinic orbit (Fig.\
\ref{fig:homoevectangents}) can develop transverse intersections. In
other words, the stable and unstable manifolds do not coincide but
rather intersect, and once they intersect, they do so an infinite
number of times creating a homoclinic tangle, as in Fig.\
\ref{fig:tangle}. The homoclinic tangle is associated with a fractal
set of periodic orbits and marks the locus of chaotic behavior.
Chaotic behavior has in fact already been found in the Kerr system for
spinning test particle motion
\cite{suzuki1997}
and in the case of spinning comparable mass black holes
\cite{levin2000,levin2003}.

Chaos may be dissipated by gravitational radiation losses
\cite{hartl2003b,hartl2005,cornish2003:2}. However,
due to the poverty of the approximation methods
in the strong-field, there is no
definitive resolution to the question of the survival versus extinction of chaos in
astrophysical systems. If chaos does survive radiative dissipation in
rapidly spinning black hole pairs, the
highly non-linear character of black hole spacetimes could be
evidenced by
the destruction of the homoclinic orbit on transition to
plunge. 

\acknowledgements

We are especially grateful to Becky Grossman for her valuable and
generous contributions to this work.  We also thank Bob Devaney for
helpful input concerning dynamical systems language.  JL and GP-G
acknowledge financial support from a Columbia University ISE
grant. This material is based in part upon work supported under a
National Science Foundation Graduate Research Fellowship.

\vfill\eject
\appendix

\section{Derivation of Action of Homoclinic Orbits}
\label{append1}

The radial action of a bound non-plunging orbit is the area enclosed
by its projection into the $r,p_r$ plane,
\begin{equation}
\label{eq:Jrdef}
  J_r \equiv \oint p_r(r) \, dr = 
  2 \int_{r_p}^{r_a} dr \frac{\sqrt{R(r)}}{\Delta}
  \quad ,
\end{equation}
where $r_p$ and $r_a$ are the periastron and apastron, respectively,
and $R(r)$ is the function (\ref{eq:Rpoly}).

For a homoclinic orbit, $r_p$ equals $r_u$, the radius of the
associated unstable circular orbit, and $r_a$ is expressible in terms
of $r_u$ alone \cite{levin2008:3}.  Additionally, $R(r)$ factors into
\begin{equation}
\label{eq:Rhcfactored}
  R(r) = (1 - E^2) r (r - r_u)^2 (r_a - r)
  \quad ,
\end{equation}
with $E$ the common energy of the homoclinic and unstable circular
orbit.  The orbit independent quantity $\Delta$ can always be
factored into
\begin{equation}
\label{eq:Deltafactored}
  \Delta = (r - r_{\sss{+}}) (r - r_{\sss{-}})
  \quad ,
\end{equation}
where $r_{\sss{\pm}} \equiv 1 \pm \sqrt{1 - a^2}$ are the outer and
inner horizons, respectively, of the central black hole.  Together,
the above allows us to write the radial action (\ref{eq:Jrdef}) of a
homoclinic orbit as
\begin{equation}
\label{eq:Jrhcdef}
  \begin{split}
    \frac{J_r^{\text{hc}}}{2 \sqrt{1 - E^2}} &=
    \int_{r_u}^{r_a} dr 
    \frac
	{(r - r_u) \sqrt{r(r_a - r)}}
	{(r - r_{\sss{+}}) (r - r_{\sss{-}})} \\
    &=
    \int_{r_u}^{r_a} dr \sqrt{\frac{r_a - r}{r}}  
    \frac{r (r - r_u)}{(r - r_{\sss{+}}) (r - r_{\sss{-}})}
  \end{split}
  \, .
\end{equation}
\ph{.}

The integral in (\ref{eq:Jrhcdef}) can be done analytically.  Under
the change of variable
\begin{equation}
\label{eq:varchange}
  \begin{split}
    u = \sqrt{\frac{r}{r_a - r}} \,, &\mrph{=}
    r = \frac{u^2}{u^2 + 1}r_a \\
    dr\,\sqrt{\frac{r_a - r}{r}} &=
    du\, \frac{2 r_a}{\lf( 1 + u^2 \rt)^2}
  \end{split}
  \quad,
\end{equation}
the factors in (\ref{eq:Jrhcdef}) become
\begin{equation}
\label{eq:Jrfactorsu}
  \begin{split}
    r_a - r &=
    \frac{r_a}{1 + u^2} \\
    r - r_u &= 
    \frac{u^2(r_a - r_u) - r_u}{1 + u^2} \\
    r - r_{\sss{+}} &= 
    \frac{u^2(r_a - r_{\sss{+}}) - r_{\sss{+}}}{1 + u^2} \\
    r - r_{\sss{-}} &=
    \frac{u^2(r_a - r_{\sss{-}}) - r_{\sss{-}}}{1 + u^2}
  \end{split}
\end{equation}
and (\ref{eq:Jrhcdef}) becomes
\begin{equation}
\label{eq:Jrhcu}
  \begin{split}
    \nquad\!
    \frac{J_r^{\text{hc}}}{2 \sqrt{1 - E^2}} &=
    \frac{r_a - r_u}
	 {\lf( r_a - r_{\sss{+}} \rt) \lf( r_a - r_{\sss{-}} \rt)}
    \times {} \\
    &\mrph{=} \int_{u_u}^{\infty} du\,
    \frac{2 r_a^2 \, u^2 \lf[ u^2 - u_u^2 \rt]}
    {
      \lf( 1 + u^2 \rt)^2
      \lf[ u^2 - u_{\sss{+}}^2 \rt]
      \lf[ u^2 - u_{\sss{-}}^2 \rt]
    }
  \end{split}
  \, ,
\end{equation}
where
\begin{alignat}{3}
\label{eq:special-u's}
    u_u^2 &\equiv \frac{r_u}{r_a - r_u}\,, \,\, &
    u_{\sss{+}}^2 &\equiv \frac{r_{\sss{+}}}{r_a - r_{\sss{+}}} \,,
    \,\, &
    u_{\sss{-}}^2 &\equiv \frac{r_{\sss{-}}}{r_a - r_{\sss{-}}}
    \quad .
\end{alignat}

The integral in (\ref{eq:Jrhcu}) decomposes by partial fractions into
\begin{equation}
\label{eq:JrwithIsandAs}
    \frac{J_r^{\text{hc}}}{2 \sqrt{1 - E^2}} =
    \lf(
    A_{1}\mathcal{I}_1 + A_{2}\mathcal{I}_2
    + A_{3}\mathcal{I}_3 + A_{4}\mathcal{I}_4
    \rt)
    \bigg\rvert^{\infty}_{u_u}
  \quad ,
\end{equation}
where the coefficients $A_i$ are
\begin{equation}
  \begin{split}
    A_{1} = r_a \,, &\quad 
    A_{2} = 2 \lf (r_u - 2 \rt) \\
    A_{3} = 
    \frac{ r_{\sss{-}} \lf( r_u - r_{\sss{-}} \rt) }
	 {\sqrt{1 - a^2}} \,,
    &\quad
    A_{4} = 
    - \frac{r_{\sss{+}} \lf( r_u - r_{\sss{+}} \rt)}{\sqrt{1 - a^2}}
  \end{split}
  \quad .
\end{equation}
and the functions $\mathcal{I}_i$ are
\begin{widetext}
\begin{subequations}
\label{eq:4uints}
\begin{align}
  \begin{split}
    \mathcal{I}_1 &\equiv \int du \, \frac{2}{\lf( 1 + u^2 \rt)^2} 
    = \frac{u}{1 + u^2} + \atan u
  \end{split}
  \label{eq:I1int}\\
  \begin{split}
    \mathcal{I}_2 &\equiv \int du \, \frac{1}{1 + u^2} 
    = \atan u
  \end{split}
  \label{eq:I2int}\\
  \begin{split}
    \mathcal{I}_3 &\equiv\ph{-} \int du 
    \, \frac{1}{u^2 - u_{\sss{-}}^2} 
    =  \half \ph{-} \sqrt{ \frac{r_a - r_{\sss{-}}}{r_{\sss{-}}} }
      \ln \lf[ \frac{u - u_{\sss{-}}}{u + u_{\sss{-}}} \rt]
  \end{split}
  \label{eq:I3int}
  \quad . \\
  \begin{split}
    \mathcal{I}_4 &\equiv \int du 
    \, \frac{1}{u^2 - u_{\sss{+}}^2} 
    = \half  \sqrt{ \frac{r_a - r_{\sss{+}}}{r_{\sss{+}}} }
      \ln \lf[ \frac{u - u_{\sss{+}}}{u + u_{\sss{+}}} \rt]    
  \end{split}
  \label{eq:I4int}
\end{align}
\end{subequations}

The right hand side of (\ref{eq:JrwithIsandAs}) is easiest to evaluate
in pieces.  The first two terms give
\begin{equation}
\label{eq:JrAI12eval}
  \begin{split}
    \lf( A_1 \mathcal{I}_1 \rt. &+ \lf. A_2 \mathcal{I}_2 \rt)
    \bigg\rvert^{\infty}_{u_u} 
=
    -r_a \frac{u_u}{1 + u_u^2}
    \lf( r_a + 2 r_u - 4 \rt) \lf( \frac{\pi}{2} - \atan u_u \rt)
    \\
    &= -\sqrt{r_u \lf( r_a - r_u \rt)}
+    \lf( r_a + 2 r_u - 4 \rt) \lf( \atan \frac{1}{u_u} \rt)
    \\
    &= -\sqrt{r_u \lf( r_a - r_u \rt)}
+ 
    2\, \frac{2E^2 - 1}{1 - E^2} 
    \atan \sqrt{ \frac{r_a - r_u}{r_u} }
  \end{split}
  \, .
\end{equation}

To go from the first to the second line in (\ref{eq:JrAI12eval}), we
have used $\atan(u) + \atan(1/u) = \pi/2$.  To get the last line, we
have used the fact that
\begin{equation}
\label{eq:raplus2ru}
  r_a + 2 r_u = \frac{2}{1 - E^2}
\end{equation}
for homoclinic orbits, which follows from equating the cubic
coefficients in equations (\ref{eq:Rpoly}) and (\ref{eq:Rhcfactored}).

The third term in (\ref{eq:JrwithIsandAs}) is
\begin{equation}
\label{eq:JrAI3eval}
  \begin{split}
    A_3 \mathcal{I}_3 \bigg\rvert^{\infty}_{u_u} &=
-\half    \frac
    {
      \lf( r_u - r_{\sss{-}} \rt)
      \sqrt{ r_{\sss{-}} \lf( r_a - r_{\sss{-}} \rt) }
    }
    {
      \sqrt{1 - a^2}
    }
    \ln \lf[ \frac{u_u - u_{\sss{-}}}{u_u + u_{\sss{-}}} \rt]
    \\
    &=\half \frac{1}{\sqrt{1 - a^2}}
    \sqrt{ \frac{R(r_{\sss{-}})}{1 - E^2} }
    \ln \lf[ \frac{u_u + u_{\sss{-}}}{u_u - u_{\sss{-}}} \rt]
    \\
    &= \frac{1}{\sqrt{1 - a^2}}
    \sqrt{ \frac{R(r_{\sss{-}})}{1 - E^2} }
    \atanh \frac{u_{\sss{-}}}{u_u}
    \\
    &= \frac{1}{\sqrt{1 - a^2}}
    \sqrt{ \frac{R(r_{\sss{-}})}{1 - E^2} }
   \atanh \sqrt
    { 
      \frac{r_{\sss{-}}}{r_a - r_{\sss{-}}}
      \frac{r_a - r_u}{r_u}
    }
  \end{split}
  \quad ,
\end{equation}
and likewise
\begin{equation}
\label{eq:JrAI4eval}
  \begin{split}
    A_4 \mathcal{I}_4 \bigg\rvert^{\infty}_{u_u} &=
    -\frac{1}{\sqrt{1 - a^2}}
    \sqrt{ \frac{R(r_{\sss{+}})}{1 - E^2} }
    \atanh \sqrt
    { 
      \frac{r_{\sss{+}}}{r_a - r_{\sss{+}}}
      \frac{r_a - r_u}{r_u}
    }
  \end{split}
  \quad .
\end{equation}

Combining (\ref{eq:JrwithIsandAs}), (\ref{eq:JrAI12eval}),
(\ref{eq:JrAI3eval}) and (\ref{eq:JrAI4eval}), we find that
\begin{equation}
  \begin{split}
  J_r^{\text{hc}} &= 2 \sqrt{1 - E^2} \times
  \Biggl\{
  -\sqrt{r_u \lf(r_a - r_u \rt)}
  + 2 \frac{2E^2 - 1}{1 - E^2}
  \atan \sqrt{ \frac{ r_a - r_u  }{r_u} }
  \Biggr\}
  \\
  &\mrph{=}
  + \frac{2}{\sqrt{1 - a^2}}
  \Biggl\{
  \sqrt{R(r_{\sss{-}})}
  \atanh \sqrt
    { 
      \frac{r_{\sss{-}}}{r_a - r_{\sss{-}}}
      \frac{r_a - r_u}{r_u}
    }
  -\sqrt{R(r_{\sss{+}})}
  \atanh \sqrt
    { 
      \frac{r_{\sss{+}}}{r_a - r_{\sss{+}}}
      \frac{r_a - r_u}{r_u}
    }
  \Biggr\}
  \end{split}
  \quad .
\end{equation}
\end{widetext}


\bibliographystyle{aip.bst}
\bibliography{hctwo}

\end{document}